%% file: iclr2026_conference.tex
\documentclass{article} % For LaTeX2e
\usepackage{float} 
\usepackage{iclr2026_conference,times}
\usepackage[most]{tcolorbox}
\usepackage{fancyvrb}
\usepackage[utf8]{inputenc} % allow utf-8 input
\usepackage[T1]{fontenc}    % use 8-bit T1 fonts
\usepackage[hyperfootnotes=false]{hyperref}       % hyperlinks
\usepackage{url}            % simple URL typesetting
\usepackage{booktabs}       % professional-quality tables
\usepackage{amsfonts}       % blackboard math symbols
\usepackage{nicefrac}       % compact symbols for 1/2, etc.
\usepackage{microtype}      % microtypography
\usepackage{xcolor}         % colors
\usepackage{tabularx}
\usepackage{makecell}
\usepackage{nicematrix}
\usepackage{placeins}
\usepackage[perpage]{footmisc} 
\input{math_commands.tex}

\usepackage{hyperref}
\usepackage{url}
\usepackage{wrapfig}

\title{ReVeal: Self-Evolving Code Agents via \\Reliable Self-Verification}

\author{%
  Yiyang Jin$^{2}$\thanks{Equal contribution. Work was done during the internship at MSRA.}\quad\quad
  Kunzhao Xu$^{3}$\footnotemark[1]\quad\quad
  Hang Li$^{1}$\quad\quad
  Xueting Han$^{1}$\thanks{Project leader; Correspondence to: Xueting Han <chrihan@microsoft.com>. The project will be open-sourced in the future.}\quad\quad \\\bf
  Yanmin Zhou$^{2}$\quad\quad
  Cheng Li$^{3}$\quad\quad
  Jing Bai$^{1}$ \\
  [6pt]
  $^{1}$Microsoft Research Asia\hspace{4pt}
  $^{2}$Tongji University\hspace{4pt}
  $^{3}$University of Science and Technology of China
}

\iclrfinalcopy % Uncomment for camera-ready version, but NOT for submission.
\begin{document}

\maketitle

\begin{abstract}
% Reinforcement learning with verifiable rewards has advanced the reasoning capabilities of large language models (LLMs). However, most methods rely solely on outcome rewards without explicitly optimizing for verification and lack meaningful verification signals from realistic environments, leading to unreliable self-verification ann ineffective test-time scaling. 
% We propose \textbf{\textit{ReVeal}}, a multi-turn \textbf{\textit{Re}}inforcement learning framework that evolves code generation with explicit self-\textbf{\textit{Ve}}rification and tool-based ev\textbf{\textit{al}}uation. 
% ReVeal explicitly optimizes verification to widen the verification-generation asymmetry, making verification a reliable driver of test-time scaling. It decomposes long-horizon reasoning into iterative generation-verification turns and introduces a Turn-Aware PPO for turn-level credit assignment, which promotes the co-evolution of generation and verification while preventing reward gaming.
% At inference, the reliable self-verification enables the model to leverage self-constructed tests and external feedback to continuously evolve code, sustaining refinement for \textbf{20+} turns despite training on only three On LiveCodeBench, and significantly improves \textit{Pass@k}, expanding the reasoning boundaries of the base model. These findings highlight the promise of ReVeal as a scalable and effective paradigm for RL training and test-time scaling, paving the way for more robust and autonomous AI agents.

Reinforcement learning with verifiable rewards (RLVR) has advanced the reasoning capabilities of large language models. However, existing methods rely solely on outcome rewards, without explicitly optimizing verification or leveraging reliable signals from realistic environments, leading to unreliable self-verification and limited test-time scaling.
To address this, we widen the verification-generation asymmetry by explicitly optimizing self-verification, making it a reliable driver of deeper test-time scaling. 
We introduce \textbf{\textit{ReVeal}}, a multi-turn \textbf{\textit{Re}}inforcement learning framework that evolves code generation through self-\textbf{\textit{Ve}}rification and tool-based ev\textbf{\textit{al}}uation. ReVeal structures long-horizon reasoning as iterative generation-verification turns and incorporates TAPO for turn-level credit assignment, fostering the co-evolution of code and test generation.
At inference, this strengthened self-verification enables the model to use self-constructed tests and tool feedback to continuously evolve code for \textbf{20+} turns on LiveCodeBench despite training on only three. It also significantly improves \textit{Pass@k}, indicating stronger exploration that expands the reasoning boundaries of the base model. These findings highlight the promise of ReVeal as a scalable paradigm for RL training and test-time scaling, paving the way for more robust and autonomous AI agents.
\end{abstract}

\begin{figure}[h]
    \centering
    \includegraphics[width=\textwidth]{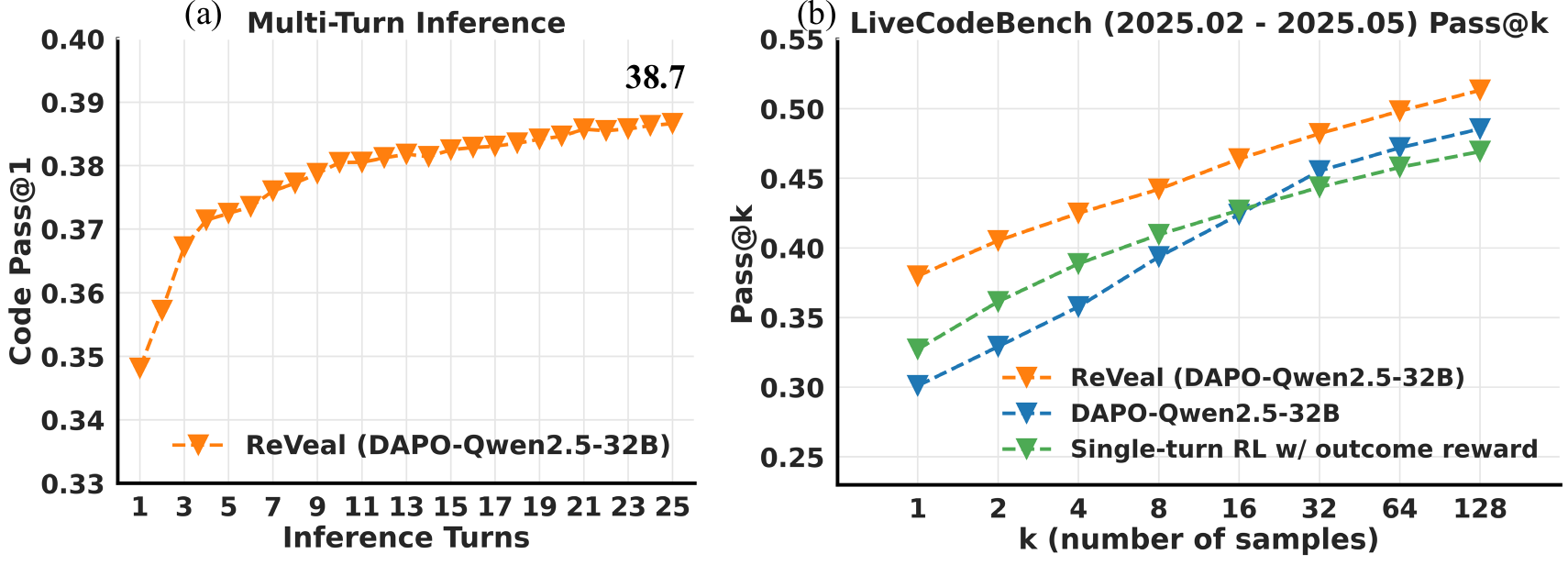}  
    \vskip -0.1in
    \caption{Performance of ReVeal on LiveCodeBench V6. (a) ReVeal enables effective test-time scaling, with Pass@1 accuracy improving from 34.8\% at turn 1 to 38.7\% at turn 25. (b) ReVeal (max\_turn=10) consistently outperforms both the base model and the RL baseline in Pass@k, expanding the base model’s reasoning boundaries, which the RL baseline fails to achieve.}
    % \caption{Performance of ReVeal on LiveCodeBench V6. (a) ReVeal enables effective test-time scaling, with Pass@1 accuracy surprisingly improving from 34.8\% at turn 1 to 38.7\% at turn 25, highlighting its strong potential to scale further with extended generation budgets. (b) ReVeal (max\_turn=10) consistently outperforms both the base model and the RL baseline in Pass@k, demonstrating its ability to expand the base model’s reasoning boundaries - a capability not achieved by the RL baseline.}
    \label{fig:eval}
    \vskip -0.1in
\end{figure}

\section{Introduction}
\label{sec:Introduction}
Reinforcement learning with verifiable rewards (RLVR) has recently shown strong potential to enhance the reasoning abilities of large language models (LLMs)~\citep{guo2025deepseek,openai_learning_to_reason}. A key factor behind this success is the emergence of reflection and self-verification, which allow models to iteratively refine their reasoning. Recent analyses identify the \emph{verification-generation asymmetry} (i.e., easier to verify than to solve) as the underlying mechanism for these improvements and a key driver of test-time scaling~\citep{wei2025verifierlaw,setlur2025e3learningexploreenables}. However, current RLVR methods rely solely on outcome rewards without explicitly optimizing verification. This leads to unreliable self-verification, where models often produce verbose, uninformative reflections or random guessing on hard problems, and limits the effectiveness of test-time scaling: prior studies show that reasoning performance plateaus once test-time compute exceeds the training horizon~\citep{setlur2025e3learningexploreenables}.

Complex problem-solving, such as competitive programming, typically requires multiple iterations of verification and revision rather than being solved in a single attempt, making accurate feedback essential to guide refinement. This highlights the need for verification-driven multi-turn reasoning. Prior work has attempted this either by training a separate critic model to assess each attempt—without leveraging tool feedback and at the cost of added inference-time complexity~\citep{xie2025teachinglanguagemodelscritique}—or by relying on execution feedback against pre-existing public tests, which are rarely available in real-world scenarios~\citep{gehring2025rlefgroundingcodellms}. As a result, these methods provide limited and non-generalizable verification, leaving self-verification unreliable and limiting sustained improvement.

% Complex problem-solving, such as challenging coding tasks, often requires multiple iterations of verification and revision rather than being solved in a single attempt. Accurate verification and feedback are therefore crucial to guide this refinement. Prior work has explored multi-turn code generation either by training a separate critic model to assess each attempt, which increases inference-time complexity~\citep{xie2025teachinglanguagemodelscritique}, or by relying on execution feedback against pre-existing public tests, which are typically unavailable in real-world scenarios~\citep{gehring2025rlefgroundingcodellms}.

% To address these limitations, we propose ReVeal, a reinforcement learning framework that explicitly optimizes verification through dense, turn-level rewards and tool-based feedback, enabling reliable self-verification and sustained multi-turn improvements.

\begin{wrapfigure}{r}{0.5\textwidth}
  \vspace{-12pt} % 图整体往上拉
  \begin{center}
    \includegraphics[width=0.5\textwidth]{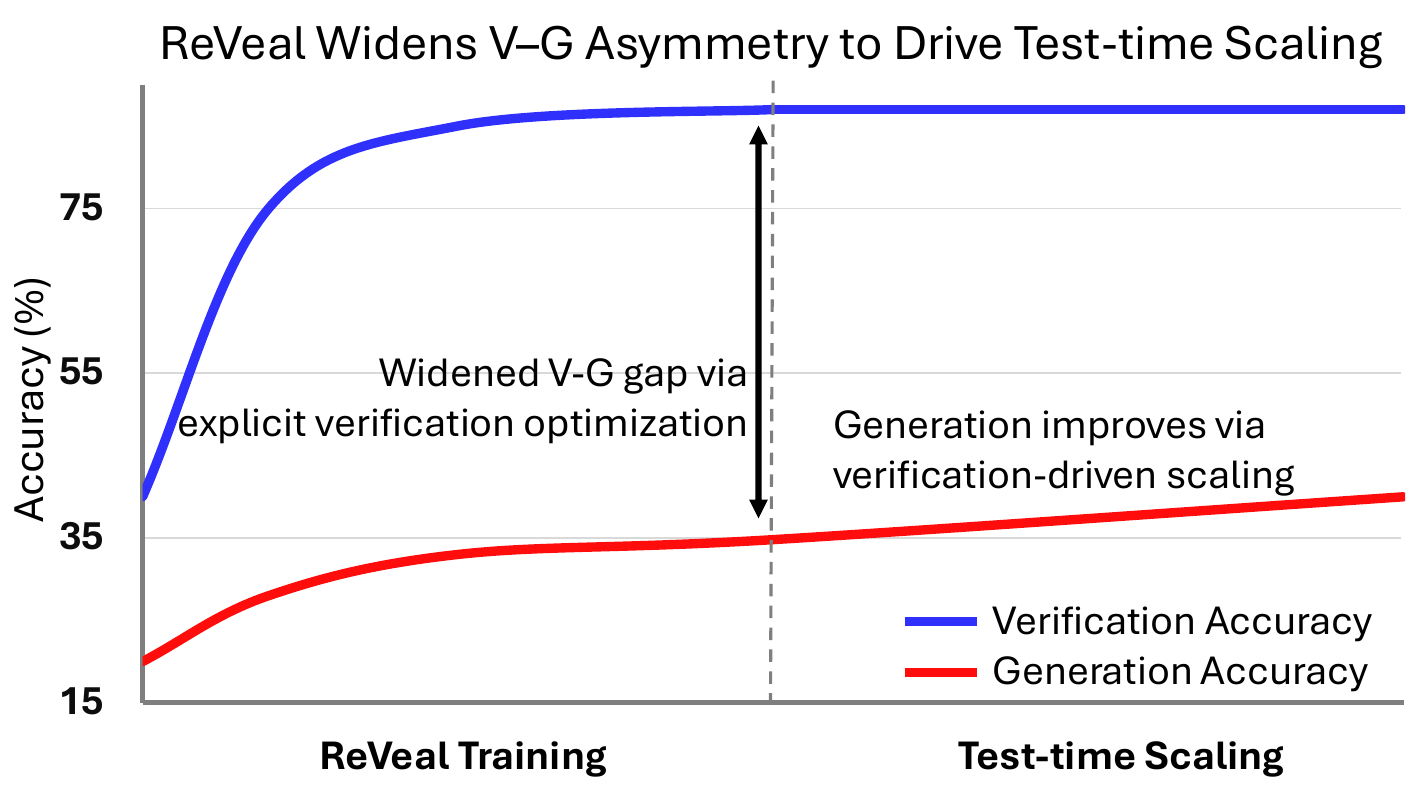}
  \end{center}
  \vspace{-12pt} % 图和caption间的距离
  \caption{ReVeal expands the V-G gap.}
  \vspace{-12pt} % caption和正文间的距离
  \label{fig_concept}
\end{wrapfigure}
To address these limitations, we propose \textbf{ReVeal}, a multi-turn RL framework that \emph{explicitly optimizes self-verification}, thereby widening the verification-generation (V-G) asymmetry and fostering co-evolution of both capabilities during training. This enables models at inference to obtain reliable verification signals from realistic environments and iteratively refine their solutions, without needing to rely on pre-existing tests. The widened V-G gap allows verification to drive sustained improvements in generation, ultimately enabling deeper test-time scaling (Figure~\ref{fig_concept}).

Concretely, ReVeal structures long-horizon reasoning into iterative generation and verification turns. At each turn, the model generates candidate code and \emph{self-verifies} its correctness by constructing test cases and invoking external tools (e.g., a Python interpreter) for execution. This closed loop yields actionable verification signals and fine-grained feedback, allowing the model to identify errors, revise strategies, and progressively refine its output across turns.
For training, we attach \emph{dense, turn-level rewards} that directly supervise both code quality and verification accuracy. To ensure robustness, ReVeal employs a \emph{Turn-Aware Policy Optimization (TAPO)} tailored for the generation-verification interplay, assigning credit at the turn granularity and preventing reward gaming (e.g., generating trivial code to hack verification rewards).
Unlike outcome-only RL methods, ReVeal makes verification itself an optimization target, turning verification signals into reliable drivers of improvement.

We evaluate ReVeal on the challenging LiveCodeBench benchmark~\citep{jain2024livecodebenchholisticcontaminationfree}. Notably, despite being trained on only three reasoning turns, ReVeal sustains continuous refinement for over 20+ inference turns, showing robust extrapolation beyond its training horizon and tackling problems previously unsolved. 
Furthermore, ReVeal significantly outperforms the base model in Pass@k by leveraging verification signals and tool feedback to guide more effective exploration, achieving an expansion of the underlying model’s reasoning boundaries that standard RL methods fail to reach.
These results validate ReVeal as not only a practical framework for self-evolving code agents, but also as a general RL paradigm for tasks with verification-generation asymmetry, where explicitly optimizing verification unlocks reliable long-horizon reasoning.

\section{Methods}
\label{sec:Approach}
\subsection{ReVeal Framework}
\subsubsection{Iterative Generation-Verification Loop}

\begin{figure}[h]
    \centering
    \includegraphics[width=\textwidth]{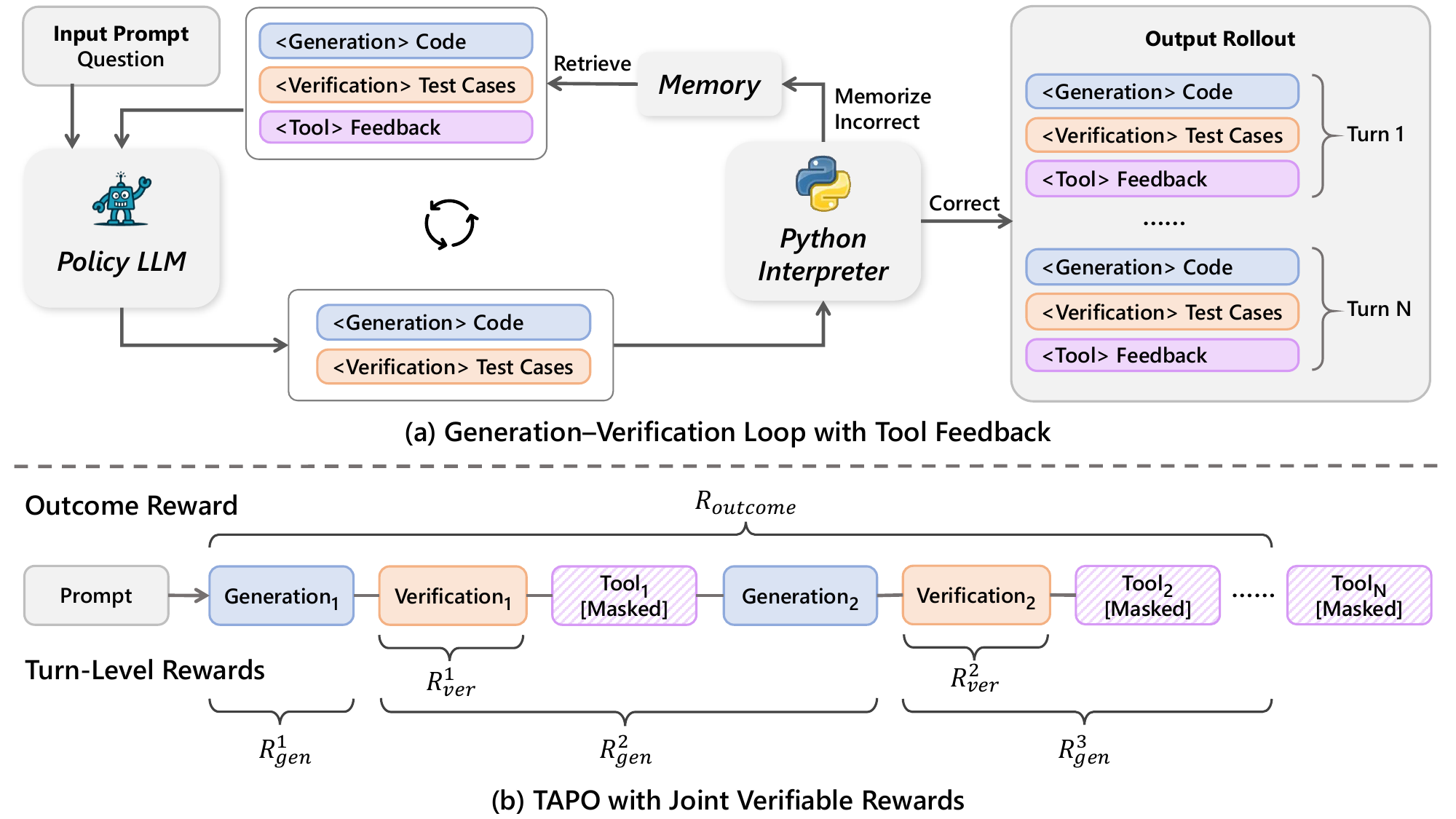}  
    \vskip -0.1in
    \caption{Illustration of ReVeal. (a) Iterative generation-verification loop with tool feedback. (b) TAPO with joint verifiable rewards: outcome, generation, and verification rewards.} 
    \label{fig:framwork}
    \vskip -0.1in
\end{figure}

ReVeal organizes long-horizon reasoning into an interleaved \emph{generation-verification} loop with tool execution feedback, where verification itself is explicitly optimized to provide reliable signals for multi-turn refinement. As illustrated in Figure~\ref{fig:framwork}, we use a single policy for both generation and verification to reduce system complexity and cost while enabling cross-capability transfer, allowing solutions and their verification strategies to co-evolve under a shared training scheme. In the code generation setting, \emph{generation} produces candidate code, whereas \emph{verification} synthesizes and executes tests to assess correctness. Fine-grained feedback from tool execution (e.g., Python interpreter) is appended to the rollout and conditions the next turn. The loop continues until a valid solution is found or a turn budget \(K\) is reached, enabling progressive refinement without external critics or predefined test cases.

% As illustrated in Figure~\ref{fig:framwork}, a single policy alternates between generating candidate solutions and constructing executable verification plans, and incorporates structured tool feedback to guides iterative revision.
% % Using one unified policy instead of separate models reduces system complexity and cost, and also enables cross-task generalization: improvements in one capability can benefit the other. Under this design, solutions and their corresponding verification strategies naturally co-evolve under our joint reward scheme. 
% In the code generation setting, \emph{generation} corresponds to producing candidate programs, while \emph{verification} means synthesizing test cases to uncover potential errors and executing them to assess correctness. Pass/fail traces and error messages from tool execution (e.g., Python interpreters) are appended to the rollout and condition the next turn. This cycle of generation, verification, and feedback continues until a valid solution is found or a maximum turn budget is reached, enabling progressive refinement across multiple reasoning turns without external critics or predefined test cases.
% To improve inference efficiency under extended multi-turn rollouts, we use a short-term memory mechanism that retains only the last three turns as context, preventing excessive context growth (detail in Appendix \ref{app: memory}).

Table~\ref{case_study} in Appendix~\ref{detailssec} illustrates a multi-turn rollout under ReVeal’s structured prompting, which decouples generation, verification, and tool feedback into distinct segments. At each turn, the policy first reasons thoroughly and explores diverse reasoning patterns freely, then emits structured outputs: executable code in \texttt{<generation-answer>} and executable tests in \texttt{<verification-answer>}. As shown in the case study, after producing candidate code the model begins verification: it hypothesizes potential failure modes and edge conditions to propose diverse test cases. The \texttt{<tool-feedback>} section then records execution results, including runtime errors, invalid test cases, as well as the expected output, actual output, and pass/fail judgment for each valid test case. Based on this feedback, the model interprets traces and error messages, diagnoses underlying causes, and adjusts both its candidate code and its verification plan in the next turn.
% This design ensures that every iteration yields executable artifacts and that raw tool outputs are transformed into fine-grained feedback that guides subsequent reasoning. 
Full prompting and feedback templates are provided in Appendix~\ref{detailssec} Tables~\ref{Prompt_Table} and \ref{feedback_mapping}.

\subsubsection{Tool-Augmented Verification} The interaction with external tools provides reliable, fine-grained supervisory signals that condition subsequent reasoning and enable systematic refinement of both code and verification strategies across turns.
More importantly, tool interaction broadens exploration during reinforcement learning by revealing concrete failure modes, steering the policy into promising regions of the search space beyond a single attempt and helping it escape local optima. Empirically (see §\ref{sec:analysis}), this yields consistently higher Pass@k than the base model.

During RL training, the \texttt{<tool-feedback>} section is excluded from the loss and used only as contextual input, which stabilizes optimization while preserving coherent multi-turn rollouts.
To ensure feedback quality during training, we adopt a filtering mechanism: model-generated test cases are executed on candidate code only if they are verified against a golden solution. This guarantees that execution traces provide legitimate supervision, thereby improving feedback precision and guiding exploration toward correct solutions. At test time, no golden reference is available; all generated test cases are executed, making verification fully autonomous. This places a strong demand on the model’s ability to generate high-quality tests. To meet this demand, ReVeal adopts a novel RL algorithm that incentivizes diverse and reliable test construction.

\subsection{Turn-Aware RL for the Generation-Verification Paradigm}
Prior RLVR methods rely on outcome-only signals to optimize an entire long reasoning trace, but this provides imprecise credit to intermediate verification and often degenerates into blind reflection. Yet one may ask: \textit{can the current paradigm fully sustain reliable verification and deeper test-time scaling?} Verification, however, is a non-trivial task: with well-designed verifiable rewards, a task can often be solved effectively. This motivates ReVeal to explicitly optimize verification with hard-to-hack rewards, which widen the verification-generation gap. At test time, this asymmetry becomes an asset: easier and more reliable verification signals can effectively guide the harder generation process to evolve over many turns.
\subsubsection{Joint Verifiable Rewards}

% While outcome-only rewards in RL have shown promise in enabling LLMs to learn self-verification, we ask: \textit{is the current paradigm fully leveraging this capability? Can denser, higher-quality rewards lead to more effective training?} To address these questions, we propose a  structured set of turn-level, verifiable rewards for both generation and verification turns. 
% Our extensive experiments demonstrate that, when guided by dense turn-level supervision, the generation-verification paradigm not only effectively optimizes both generation and verification capabilities but also enables effective test-time scaling, supporting continuous performance improvement beyond the original training regime.

To jointly train generation and verification, ReVeal decomposes the reward into three complementary components (Figure~\ref{fig:framwork}(b)): an \emph{outcome reward} supervising the final solution, a \emph{generation reward} capturing improvements across generation turns, and a \emph{verification reward} evaluating the quality of generated tests. This design naturally links the two roles in a co-evolutionary loop.

\noindent\textbf{Outcome reward.}
The outcome reward shapes the entire reasoning chain by the final solution quality:
\begin{equation}
r_{\mathrm{outcome}} = r_{\mathrm{format}} + r_{\mathrm{passrate}},
\end{equation}
where the format reward $r_{\mathrm{format}}$ ensures that the model produces well-formed generation and verification blocks,
\begin{equation}
r_{\mathrm{format}} = 
\begin{cases}
1, & \text{if the output format is correct},\\
-1, & \text{otherwise},
\end{cases}
\end{equation}
and $r_{\mathrm{passrate}} = 5 \times \mathit{passrate}$ measures final code accuracy with $\mathit{passrate}\in[0,1]$, giving $r_{\mathrm{outcome}}\in[-1,6]$.

\noindent\textbf{Generation reward.}
For each generation turn $k$ (odd), we compute the pass rate $r_{\mathrm{passrate}}^k$ of the code produced and define:
\[
r_{\mathrm{gen}}^k =
\begin{cases}
r_{\mathrm{passrate}}^1, & k = 1,\\[6pt]
\mathit{abs}\cdot r_{\mathrm{passrate}}^k + \mathit{imp}\cdot \bigl(r_{\mathrm{passrate}}^k - r_{\mathrm{passrate}}^{k-2}\bigr), & k \ge 3,
\end{cases}
\]
where $\mathit{abs}$ and $\mathit{imp}$ weight absolute accuracy and iterative improvement. We set $\mathit{abs}=0,\ \mathit{imp}=1$ so that the reward encourages real improvements in code accuracy across turns.

\noindent\textbf{Verification reward.}
For each verification turn $k$ (even), we reward the proportion of generated tests that succeed when executed on a golden code:
\begin{equation}
r_{\mathrm{ver}}^k = \frac{\#\{\text{test cases in turn }k\text{ that pass}\}}
                           {\#\{\text{test cases generated in turn }k\}}.
\end{equation}

\subsubsection{Turn-Aware Policy Optimization}

\paragraph{Preliminaries.}
Our algorithm builds on the Proximal Policy Optimization (PPO) framework~\citep{schulman2017proximalpolicyoptimizationalgorithms}, an on-policy actor-critic method that optimizes a clipped surrogate objective for stable updates. PPO typically estimates token-level advantages using Generalized Advantage Estimation (GAE)~\citep{schulman2018highdimensionalcontinuouscontrolusing}:
\begin{equation}
\hat{A}_t^{\mathrm{GAE}(\gamma,\lambda)} 
= \sum_{l=0}^{\infty} (\gamma\lambda)^l \bigl(r_{t+l} + \gamma V_{t+l+1} - V_{t+l}\bigr),
\end{equation}
where $\gamma\in[0,1]$ is the discount factor and $\lambda\in[0,1]$ controls the bias-variance trade-off.  

\paragraph{Turn-Aware Policy Optimization.}
Building on our structured reward design, we introduce \emph{Turn-Aware Policy Optimization} (TAPO), which preserves the PPO actor-critic framework but replaces GAE-based advantages with a \emph{turn-aware} return. TAPO leverages the critic to efficiently bootstrap from both token-level Monte Carlo returns and turn-level returns, enabling stable learning across these two reward granularities.
% TAPO combines token-level Monte Carlo returns with turn-level returns, enabling stable credit assignment across two granularities.  

1. \textbf{Token-level return.}  
We set $\lambda=1$ and $\gamma=1$ (pure Monte Carlo). For token step $t$ with final step $T$:  
\begin{equation}
R_t = \sum_{l=0}^{T-t} r_{t+l} \;=\; r_t + R_{t+1}, 
\quad R_{T+1}=0.
\end{equation}

2. \textbf{Turn-level return.}  
To mitigate adversarial reward gaming (e.g., generating trivial code that hacks the verification reward), we introduce a \emph{turn-level} return tailored to the generation-verification interplay. Specifically, (i) each generation reward is assigned both to its own generation turn and to the immediately preceding verification turn, and (ii) each verification reward is confined strictly to its own verification turn. This design prevents reward hacking by ensuring that generation turns are rewarded solely based on code quality, rather than verification success. Let $\{t_1, \dots, t_K\}$ denote the token indices at which each turn ends  (alternating generation and verification), and define:
\begin{equation}
R^{\mathrm{turn}}(t_k) =
\begin{cases}
r_{\mathrm{gen}}^k, 
& \text{if turn }k\text{ is generation},\\[4pt]
r_{\mathrm{ver}}^k + R^{\mathrm{turn}}(t_{k+1}), 
& \text{if turn }k\text{ is verification},
\end{cases}
\quad
R^{\mathrm{turn}}(t_{K+1}) = 0.
\end{equation}
For token $t$, let $\tau(t)=\min\{t_k \mid t_k \ge t\}$ and define
\begin{equation}
R^{\mathrm{turn}}_t =
\begin{cases}
R^{\mathrm{turn}}\bigl(\tau(t)\bigr), & \text{if }\tau(t)\text{ exists},\\
0, & \text{otherwise}.
\end{cases}
\end{equation}

3. \textbf{Turn-aware return.}  
The final return combines the two levels:
\begin{equation}
\widetilde R_t = R_t + R^{\mathrm{turn}}_t,
\qquad
A_t = \widetilde R_t - V_t,
\end{equation}
where $V_t$ is the critic model’s estimate at step $t$. These advantages $A_t$ then replace the standard GAE estimates in the PPO objective, completing the TAPO update.  

% \noindent\textbf{Interaction and co-evolution.}
% These rewards work in concert: outcome rewards align the process with final correctness, and turn-level signals provide dense supervision for progressive refinement. In addition, the interplay between tests and code creates a feedback loop: stronger tests expose errors that drive code improvements, which are then reinforced by the generation reward; meanwhile, improved code raises the bar for verification, pushing the model to design richer and more challenging tests. Through this loop, code and tests co-evolve, driving steady accuracy improvements during RL training.
\paragraph{Discussion.}
TAPO provides sharper supervision than outcome-only methods by explicitly assigning credit at both token and turn levels. It integrates outcome rewards, which keep the process aligned with final correctness, and turn-level signals, which provide dense supervision for progressive refinement. This structure establishes a feedback loop: stronger tests expose errors that drive code improvements, which are then reinforced by the generation reward, while improved code raises the bar for verification, pushing the model to generate richer and more challenging tests. By design, TAPO prevents reward gaming and turns this loop into stable co-evolution of code and tests. Crucially, TAPO is a \emph{general} credit-assignment algorithm, applicable to any reasoning task with verifiable rewards for both generation and verification.

\section{Experiments}
\label{sec:Experiments}
\subsection{Settings}
\paragraph{Dataset} We construct our training dataset from TACO \citep{li2023tacotopicsalgorithmiccode}, a large-scale corpus comprising  26,443 algorithmic programming problems sourced from competitive programming platforms such as LeetCode \citep{leetcode} and Codeforces \citep{codeforces}. Each problem consists of a natural language description, golden solutions, and multiple test cases.

To address noise in the raw dataset, we first filter out problems containing unsupported content types, specifically those tagged with interactive or image elements. To ensure testability and correctness, we process two types of test-case formats, function-based tests and standard input/output tests, into a unified structure compatible with our code execution environment. We then execute each test case against the first available golden solution in our execution environment. Problems where the golden code fails to pass all associated test cases are discarded. After preprocessing, we retain a high-quality dataset of 11,151 problems for training and 509 problems for testing.

% \paragraph{Models and Training Details}
% We use Qwen2.5-32B-Instruct~\citep{qwen2025qwen25technicalreport} as our main backbone. However, when applying RL to this model using only code-related data, we did not observe the spontaneous emergence of long chain-of-thought capabilities with reflection, consistent with the findings of SRPO~\citep{zhang2025srpocrossdomainimplementationlargescale}. To address this, we adopt DAPO-Qwen-32B~\citep{yu2025dapoopensourcellmreinforcement} as our base model, which is reinforced with mathematical data, and we continue RL training on code datasets to adapt its reasoning capabilities to coding tasks. Our models are trained using the Verl~\citep{sheng2024hybridflow} framework on 8/16 AMD Mi300x GPUs. The RL training process follows the hyperparameter settings listed in Table \ref{RL_Training_Hyperparameters}.We set the maximum turns to 3 during RL training.
\paragraph{Models and Training Details}
We adopt DAPO-Qwen-32B~\citep{yu2025dapoopensourcellmreinforcement} as our base model, which is reinforced with mathematical data, and we continue RL training on code datasets to adapt its reasoning capabilities to coding tasks. Our models are trained using Verl~\citep{sheng2024hybridflow} framework on 8/16 AMD Mi300x GPUs. The RL training process follows the hyperparameter settings listed in Table \ref{RL_Training_Hyperparameters}. We set maximum turns to 3 during RL training.

\paragraph{Evaluation} We evaluate ReVeal on two code generation benchmarks: LiveCodeBench (LCB) V6 (2025.02-2025.05) \citep{jain2024livecodebenchholisticcontaminationfree} and CodeContests \citep{li2022competition}. The evaluation process follows the hyperparameter configuration specified in Table \ref{RL_Evaluation_Hyperparameters}. Although training is performed with a maximum of 3 turns, we evaluate the model under extended turn settings (8 and 25 turns) to assess its generalization to longer reasoning horizons and test-time scaling performance.

We use Pass@1 to measure the success rate of the model's final code solutions. To evaluate the model's verification and self-correction capabilities, we introduce two additional metrics: $\Delta_{\uparrow}$ denotes the fraction of initially incorrect solutions that become correct after revision, and $\Delta_{\downarrow}$ denotes the fraction of initially correct solutions that become incorrect after revision. In line with recent work~\citep{yue2025doesreinforcementlearningreally}, we use Pass@k up to \(k=128\) to assess whether ReVeal can push the reasoning boundaries beyond the base model, with at most 10 generation-verification turns per example.

\paragraph{Memory Mechanism for Context Management}
To improve inference efficiency under extended multi-turn rollouts, we use a short-term memory mechanism that retains only the last three turns as context, which prevents excessive context growth without hurting accuracy (details in Appendix \ref{app: memory}).

\paragraph{Code Execution Tool} We use Code Judge\footnote{\url{https://github.com/0xWJ/code-judge}} as our code execution environment. Code Judge supports both function-based and standard input-output test case formats through a consistent interface. Designed for scalability and robustness, it enables efficient long-batch execution through multi-processing and provides reliable code evaluation.
\paragraph{Baselines}
We compare ReVeal against the following baselines: (1) \textit{Base}: base models without code-specific RL training; (2) \textit{CTRL~\citep{xie2025teachinglanguagemodelscritique} + Qwen2.5-Coder-32B-Instruct}: five-turn critic-revision with a dedicated critic model; results cited from the original paper (evaluated on LCB 24.08-24.11); (3) \textit{Single-turn RL with outcome reward}: RL with outcome-only rewards under standard <think>-<answer> prompting template without any external tool calls. %; (4) \textit{DeepSeek-R1-Zero-Qwen-32B}: a strong RL-zero baseline trained on both math and code data without any supervised fine-tuning.

\begin{table*}[t]
    \centering
    \caption{Performance comparison of ReVeal with baseline methods on LiveCodeBench V6 and CodeContests. Pass@1 indicates the success rate; $\Delta_{\uparrow}$ and $\Delta_{\downarrow}$ represent the percentages of incorrect solutions corrected and correct solutions degraded after revision, respectively.}
    \begin{NiceTabularX}{\textwidth}{lX[c]X[c]X[c]X[c]X[c]X[c]}
    \toprule
    % Model column spans 2 rows. LiveCodeBench spans the 3 X-columns.
    \textbf{Model} & \multicolumn{3}{c}{\textbf{LiveCodeBench V6}} & \multicolumn{3}{c}{\textbf{CodeContests}} \\
    \cmidrule(lr){2-7} % Rule under LiveCodeBench, spanning columns 2 to 4
    % Sub-headers for the 3 data columns
    & \textbf{Pass@1} & $\Delta_{\uparrow}$ & $\Delta_{\downarrow}$ & \textbf{Pass@1} & $\Delta_{\uparrow}$ & $\Delta_{\downarrow}$\\
    \midrule
    \multicolumn{7}{c}{\rowcolor{gray!15}\textit{Existing Baselines}} \\
    Qwen2.5-32B-Instruct            & 24.8  & -     & -  & 13.3  & -     & -    \\
    DAPO-Qwen-32B & 31.1 & - & - & 18.5 & - & - \\
    Qwen2.5-Coder-32B-Instruct      & 29.5    & -     & -  & 14.6    & -     & -    \\
    \phantom{1ex}w/ critic$\times$5 Qwen2.5-Coder & 29.6 & 2.14   & 3.04 & -    & -     & -   \\
    \phantom{1ex}w/ critic$\times$5 GPT-4o         & 32.9 & 4.82  & 2.50 & -    & -     & -  \\
    \phantom{1ex}w/ critic$\times$5 CTRL           & 33.4 & 3.75  & 0.89 & -    & -     & -  \\
    \midrule
    \multicolumn{7}{c}{\rowcolor{gray!15}\textit{RL based on DAPO-Qwen-32B}} \\
    Single-turn RL & 32.8  & -   & - & 21.0  & -   & -  \\
    ReVeal$\times$25 & \textbf{38.7}  & \textbf{7.50}  & \textbf{0.0} & \textbf{33.6}  & \textbf{15.69}  & \textbf{0.0}  \\
    \midrule
    \multicolumn{7}{c}{\rowcolor{gray!15}\textit{Ablation Study: TAPO with Joint Verifiable Rewards}} \\
    ReVeal$\times$8 w/ outcome reward & 36.1    & 4.69   &  1.32 & 27.4    & 9.24   &  2.36  \\
    ReVeal$\times$8 w/ TAPO with joint rewards & 37.7  & 5.62  & 0.0 & 30.4  & 12.30  & 0.0   \\
    \bottomrule
    \end{NiceTabularX}
    \label{table:main}
\end{table*}

\subsection{Main Results}
Table~\ref{table:main} shows that single-turn RL (outcome-only, no explicit optimization of self-verification or tool use) improves Pass@1 over the base models. ReVeal goes further by explicitly optimizing verification and enabling deeper inference; it surpasses the single-turn RL baseline by a wide margin. In addition to deeper-turn gains, ReVeal also achieves higher Pass@1 at turn 1 (34.8\%) than the single-turn RL baseline under the same inference budget on LCB V6, indicating that multi-turn training (3 turns) transfers exploration benefits into a stronger policy and that increasing training depth may further amplify gains.

ReVeal significantly outperforms critic-based methods such as CTRL. While critic models tailored for code tasks can be paired with policy models for multi-turn critique and revision, ReVeal employs a single policy model that self-verifies and iteratively refines its own outputs, yet achieves superior results, highlighting the benefit of jointly optimizing generation and verification. Specifically, ReVeal attains larger correction rates with near-zero degradation, demonstrating highly robust and reliable capabilities in self-verification, critique, and revision. (CTRL numbers are cited from earlier LCB version; see Table~\ref{table:main3} for a V5 comparison on Qwen2.5-32B-Instruct.)

Ablation studies confirm the benefit of TAPO with joint verifiable rewards: at the same turn budget it yields higher Pass@1, increases $\Delta_{\uparrow}$, and suppresses $\Delta_{\downarrow}$ compared to outcome-only training. In contrast, outcome-only rewards exhibit higher $\Delta_{\downarrow}$, indicating that insufficiently optimized verification can drive incorrect revisions.

\paragraph{More Experiments} 
To validate effectiveness and scalability across models, we evaluate ReVeal on another base model, Qwen2.5-32B-Instruct. Detailed results are provided in Appendix~\ref{sec:qwen32b_more} Table~\ref{table:main3} and Figure~\ref{fig:experiment2}. ReVeal outperforms single-turn RL baseline by 4.1\%, which demonstrates the effectiveness of ReVeal. To demonstrate that the performance improvements are statistically significant, we repeated the experiment 8 times and reported the $mean\pm std$ in the Appendix~\ref{sec:qwen32b_more} Table~\ref{table:main4}. Across models of varying capability, ReVeal remains effective. With the stronger DAPO-Qwen-32B backbone, ReVeal unlocks greater headroom: accuracy continues to improve with deeper inference turns and surpasses outcome-only RL by a wider margin. This underscores ReVeal’s potential on stronger backbones.

To verify that the performance improvements originate from explicit optimization rather than mere tool use and prompting, we evaluate a \emph{prompt-only ReVeal} baseline using the same tools and prompt, showing that ReVeal significantly outperforms this baseline across all turn budgets. Without explicit optimization for verification, the prompt-only variant fails to sustain effective deep multi-turn refinement. See Appendix~\ref{promptsec} Table~\ref{table:main2} for details.

% To verify that the performance improvements originate from explicit optimization rather than mere tool use and prompting, we conduct comparative experiments with ReAct-style prompting \citep{yao2023reactsynergizingreasoningacting}, showing that ReVeal significantly outperforms ReAct across all turn budgets. Without explicit optimization for verification, ReAct fails to sustain effective deep multi-turn refinement. See Table~\ref{table:main2} for details.

\subsection{Analysis}
\label{sec:analysis}
\paragraph{ReVeal Enables Test-time Scaling into Deeper Inference Regimes.}
As shown in Figure~\ref{fig:eval}(a), ReVeal enables effective test-time scaling through iterative generation and verification. Although the model is trained with a maximum of three reasoning turns, it continues to improve its solutions when more turns are allowed at inference time, leading to progressively higher code accuracy. For instance, Pass@1 increases from 34.8\% at turn 1 to 36.7\% at turn 3, and further rises to 38.7\% by turn 25 on LiveCodeBench. 
% It is worth noting that we believe the model's performance at later turns has not yet saturated, the model has the potential to scale further when more generation budget are allowed.
This compellingly demonstrates how reliable self-verification and iterative environment feedback can enable compute scaling into deeper inference regimes, allowing ReVeal to solve previously intractable problems and evolve novel solutions. 
As a result, ReVeal supports self-improvement beyond the training horizon, enabling strong generalization in long-horizon reasoning during inference. Furthermore, these newly discovered solutions can be distilled back into the code LLM to further enhance its reasoning capabilities through continued training.

% This demonstrates that ReVeal's verification mechanism provides consistent and precise feedback across turns, allowing the model to iteratively identify and correct its own errors.

\paragraph{ReVeal Pushes Beyond the Reasoning Boundaries of the Base Model.}
We compare DAPO-Qwen-32B and single-turn RL baseline with ReVeal using Pass@k metrics on LiveCodeBench. As shown in Figure~\ref{fig:eval}(b), the RL baseline outperforms the base model when k < 32, but its performance gain gradually diminishes as k increases. In contrast, ReVeal consistently outperforms both the base model and the RL baseline across all k values from 1 to 128, demonstrating its ability to surpass the reasoning boundaries beyond the base model. We attribute this improvement to ReVeal's verification-driven exploration: tool-assisted verification provides targeted, execution-based feedback and precise judgments that guide the model to explore better solutions more effectively. With this enhanced exploration capability, the model continually self-evolves and grows beyond its initial reasoning capability during RL training. We believe this approach offers a promising path towards developing self-evolving agents with stronger reasoning capabilities. 

\begin{figure*}[!t]
    \centering
    \includegraphics[width=\textwidth]{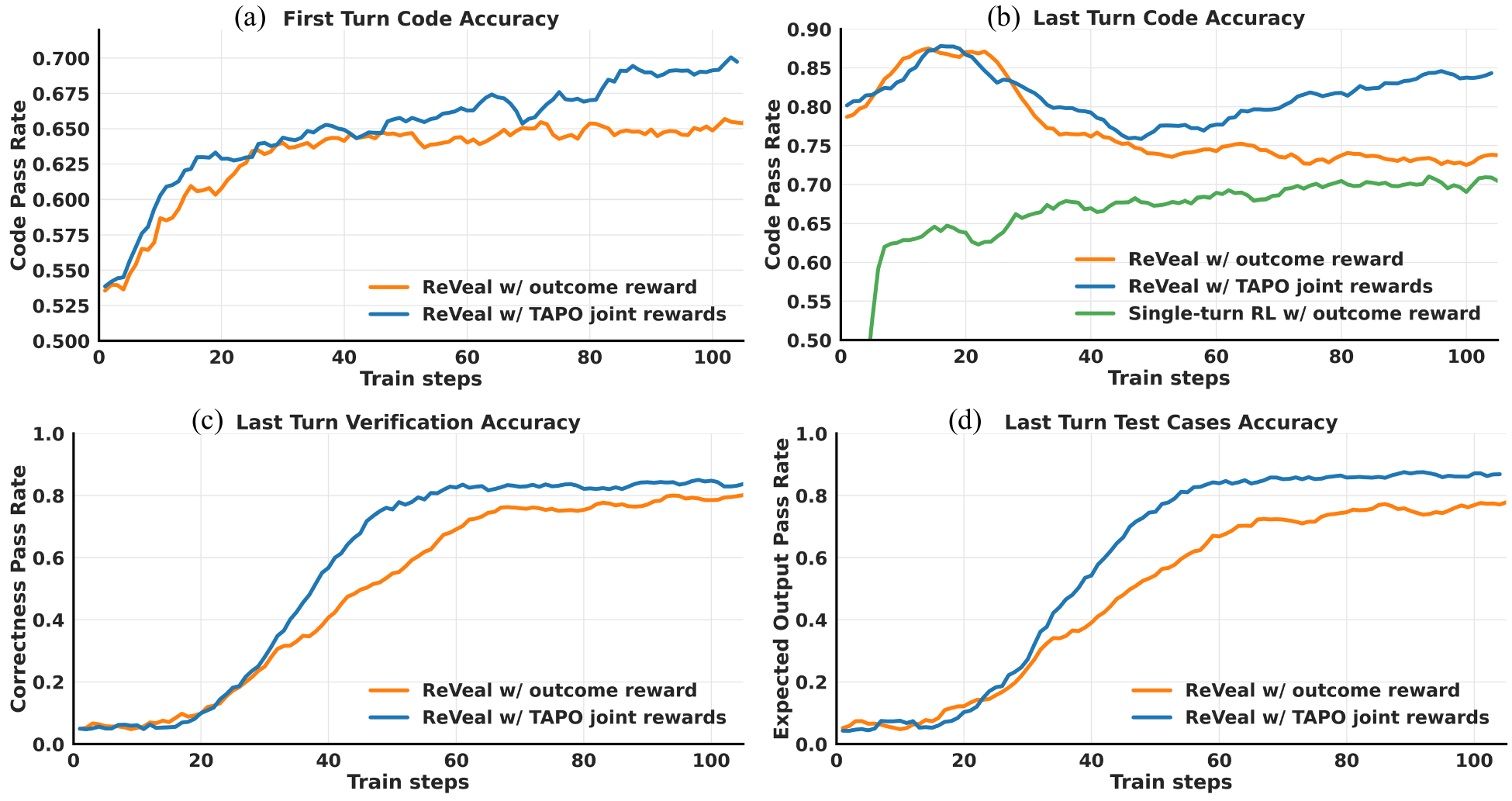}  
    \vskip -0.1in
    % \caption{Comparison of code accuracy, test case accuracy, and response length across training for ReVeal with turn-level rewards, ReVeal with outcome-only rewards, and single-turn RL without tool integration.} 
    \caption{Training curves of (a) first-turn code accuracy, (b) last-turn code accuracy, (c) last-turn verification accuracy on filtered correct test cases, and (d) last-turn test-case accuracy for three methods. \textbf{Note on (b):} the dip before step~40 is due to expanded evaluation coverage: as format score reaches 0.9 around step~40, more problems enter the evaluation set, temporarily lowering accuracy.}
    \label{fig:experiment}
    \vskip -0.1in
\end{figure*}

\paragraph{ReVeal Co-evolves the Model's Generation and Verification Capabilities.} 
Figure~\ref{fig:experiment}(b) and (d) illustrate the co-evolution of the model's code and test case generation capabilities. As shown in Figure~\ref{fig:experiment}(b), once the format is learned, final code accuracy steadily improves during training and significantly surpasses the single-turn RL baseline. Moreover, comparing Figure~\ref{fig:experiment}(a) and (b) reveals that final solutions consistently outperform those generated at turn 1, with the performance gap widening over time. This trend indicates that as the model's verification ability strengthens, multi-turn refinement enables the exploration of better solutions, progressively enhancing its capacity to generate and refine code. After the format is learned, test-case accuracy rises substantially from about 50\% at step 40 to nearly 88\%, as shown in Figure~\ref{fig:experiment}(d). Additionally, for correctly generated test cases, the model achieves over 85\% accuracy in judging code correctness (Figure~\ref{fig:experiment}(c)), indicating that the generated test cases have high error-detection coverage beyond high accuracy. This demonstrates that during inference, the model can reliably generate valid test cases and effectively leverage tools to produce accurate verification signals, which are critical for continuous improvements in code quality. These results provide strong evidence that ReVeal jointly and effectively optimizes both generation and verification, enabling the model to evolve its reasoning capabilities throughout training.

\paragraph{The Effectiveness of TAPO with Joint Verifiable Rewards.} 
As shown in Table~\ref{table:main}, TAPO with joint rewards further enhances multi-turn performance compared to relying solely on outcome rewards. Figure~\ref{fig:experiment}(a,b) show TAPO with joint rewards achieves more stable and consistent per-turn code gains, and Figure~\ref{fig:experiment}(c,d) show it achieves stronger verification (error-detection) capability and higher test-case accuracy, indicating that explicitly optimizing verification yields higher-quality tests and more effective reasoning in the code generation task.
These benefits are amplified in longer-sequence and harder verification scenarios. On the stronger DAPO-Qwen-32B backbone with longer chains (\textasciitilde5k tokens), dense turn-level supervision yields larger gains than on Qwen2.5-32B-Instruct with much shorter chains (\textasciitilde1.8k tokens; see Table~\ref{table:main3} and Figure~\ref{fig:experiment2}). This is because outcome-only signals are too coarse for extremely long chains, providing imprecise credit to intermediate verification steps. Furthermore, in more challenging verification scenarios, such fine-grained supervision becomes increasingly essential, offering richer learning signals to enhance the model’s verification capabilities.
% move to appendix
% \paragraph{The Effectiveness of Short-Term Memory.} 
% As shown in Figure~\ref{fig:experiment3}, we compare models with and without memory integration. The baseline model without memory provides complete historical information from all previous turns directly to the model, maintaining full contextual details throughout the interaction sequence. However, experimental results reveal critical limitations of this approach. While both configurations show similar performance improvements in early interaction rounds, the no-memory baseline exhibits performance saturation at deeper interaction turns due to output length constraints imposed by the model's context window. In contrast, the memory-augmented system demonstrates sustained performance gains throughout extended interaction sequences. Specifically, at 20+ interaction turns, the memory-enabled model continues to achieve incremental improvements, while the baseline's performance plateaus and occasionally degrades due to context overflow issues. This sustained improvement capability highlights the memory mechanism's effectiveness in enabling continuous learning and adaptation within computational constraints.

\section{Related Work}
\subsection{Tool-Augmented Reasoning}
Tool-integrated reasoning enables large language models (LLMs) to leverage external tools, such as search engines or code interpreters, to overcome inherent limitations in domain knowledge and mathematical operations. Early approaches demonstrated the benefits of tool integration via prompt engineering~\citep{yao2023reactsynergizingreasoningacting,chen2023programthoughtspromptingdisentangling,shinn2023reflexion} and supervised fine-tuning~\citep{gou2024toratoolintegratedreasoningagent}. ReAct~\citep{yao2023reactsynergizingreasoningacting} and Reflexion~\citep{shinn2023reflexion} interleave reasoning with actions or self-critique under tool access to refine solutions; however, their heuristic prompting often yields inaccurate tests that cause both false negatives and positives, leading to unsustained gains. More recently, multi-turn RL has been adopted to further enhance this capability on various reasoning tasks~\citep{jin2025searchr1trainingllmsreason, feng2025retoolreinforcementlearningstrategic,li2025torlscalingtoolintegratedrl}. For example, Search-R1~\citep{jin2025searchr1trainingllmsreason} incorporates multi-turn interactions with a search engine to retrieve relevant contextual information during RL training. ReTool~\citep{feng2025retoolreinforcementlearningstrategic} and ToRL~\citep{li2025torlscalingtoolintegratedrl} enable multi-turn code execution to support mathematical reasoning. Building on the promising potential of tool-integrated RL, Agent-R1~\citep{Agent-R1} introduces an open-source RL framework capable of supporting multi-turn, customizable tool invocations.

Despite their effectiveness, most tool-augmented RL methods are predominantly outcome-driven: they rely on task success or failure as the sole training signal and do not explicitly optimize verification or assign credit across turns. Likewise, prompt-only agents lack turn-level, verifiable supervision, which can make self-verification unreliable on harder problems and limit sustained test-time improvement. Unlike prior tool-augmented work, ReVeal treats verification itself as a first-class optimization target alongside generation, and introduces Turn-Aware Policy Optimization rewards (TAPO) to provide fine-grained credit to both generation and verification turns. Our approach and prior tool-augmented methods are orthogonal and complementary, and can be naturally combined to further enhance reasoning capability.
% Despite their effectiveness, these approaches are predominantly outcome-driven, rely solely on task success or failure as the training signal, without incorporating explicit verification mechanisms or fine-grained verification feedback. To address this limitation, we propose ReVeal, a multi-turn RL framework that enhances code generation through explicit self-verification, enabled by integration with external tools. This design encourages the model not only to generate solutions, but also to assess their correctness in a structured manner during training.

\subsection{Self-verification of LLMs}
Enabling LLMs to iteratively refine their outputs is critical for enhancing their reasoning capabilities. However, LLMs typically lack reliable self-judgment~\citep{huang2024largelanguagemodelsselfcorrect}. One common solution is to introduce a separate critic model to verify the output of the policy model~\citep{zhang2025generativeverifiersrewardmodeling,xie2025teachinglanguagemodelscritique}. For example, CTRL~\citep{xie2025teachinglanguagemodelscritique} uses RL to train a critic model for code completion tasks. Although effective, these approaches incur the cost and complexity of maintaining and coordinating two distinct models.

An alternative strategy is to enable one single model to generate outputs and self-verify them. In mathematical reasoning, ~\citep{xiong2025selfrewardingcorrectionmathematicalreasoning} synthesizes long chains of thought that incorporate "self-reward" and "self-correction" signals as seed data for supervised fine-tuning, and then further enhances this ability via RL. In the code domain, execution feedback effectively verifies code correctness and provides useful information for fixing errors. RLEF~\citep{gehring2025rlefgroundingcodellms} performs multi-turn code generation and verification with an integrated code execution tool; however, it depends on publicly available test cases, limiting its applicability.

In contrast, ReVeal advances self-verification by having the model generate its own high-quality test cases on the fly. By explicitly crafting and executing these tests, ReVeal eliminates the dependency on pre-existing test suites and improves applicability to real-world software systems.

\section{Conclusion}
\label{sec:Conclusion}
We presented ReVeal, a multi-turn reinforcement learning (RL) framework that makes verification a first-class optimization target alongside generation and organizes long reasoning chains into iterative generation-verification turns with tool feedback. Using TAPO with joint verifiable rewards, ReVeal equips LLMs with strong verification capabilities and demonstrates the surprising power of enabling code LLMs to self-evolve—both during RL training, where it pushes boundaries beyond the base model, and at test time, where multi-turn generation and verification continually refine outputs, even up to 20+ inference turns. This compellingly demonstrates that ReVeal can enable compute scaling into deeper inference regimes, allowing it to solve previously intractable problems and evolve novel solutions. Furthermore, these newly discovered solutions can be distilled back into the code LLM to further enhance its reasoning capabilities through continued training.

Although we demonstrate ReVeal on code tasks, its general concept of generation-verification, TAPO, and turn-level reward design can be applied to any domain with verifiable rewards for both generation and verification and that exhibits verification asymmetry, offering a promising blueprint for future advances in self-improving, more robust, and autonomous AI agents.

% We presented \textit{ReVeal}, a multi-turn reinforcement learning (RL) framework that enhances the reasoning capabilities of large language models (LLMs) through an interleaved generation-verification paradigm. By performing self-verification and integrating interactive environment feedback, this design facilitates iterative refinement over multiple reasoning turns. ReVeal equips LLMs with strong verification capabilities and demonstrates the surprising power of enabling code LLMs to self-evolve - both during RL training, where it pushes the boundaries beyond the base model, and at test time, where multi-turn generation and verification continually refine its outputs even up to 20+ inference turns. This compellingly demonstrates ReVeal can enable compute scaling into deeper inference regimes, allowing ReVeal to solve previously intractable problems and evolve novel solutions. Furthermore, these newly discovered solutions can be distilled back into the code LLM to further enhance its reasoning capabilities through continued training.
% Although ReVeal is showcased on code generation tasks, its general concept of generation and verification, and TAPO can be applied to other domains with verifiable solutions and verifications, offering a promising blueprint for future advances in self-improving, more robust and autonomous AI agents.
% \bibliographystyle{rusnat}
\newpage
\bibliography{iclr2026_conference}
\bibliographystyle{iclr2026_conference}
%%%%%%%%%%%%%%%%%%%%%%%%%%%%%%%%%%%%%%%%%%%%%%%%%%%%%%%%%%%%
\newpage
\appendix
\section{The Use of Large Language Models}
In preparing this manuscript, we used a large language model (LLM) solely for polishing the writing style and improving the clarity of the manuscript. The LLM was not used for generating research ideas, designing experiments, conducting analyses, or deriving results. All scientific contributions, including the conceptualization, methodology, experiments, and conclusions, were developed entirely by the authors.

\section{Implementation Details}
\label{detailssec}
\subsection{Hyperparameters}
Table \ref{RL_Training_Hyperparameters} and Table \ref{RL_Evaluation_Hyperparameters} show the detailed hyperparameters we use during training and evaluation.

% \vspace{-1em}
\begin{table}[H]
\centering
\caption{RL Training Hyperparameters for ReVeal-Qwen2.5-32B-Instruct and ReVeal-DAPO-Qwen-32B}
\label{RL_Training_Hyperparameters}
\begin{NiceTabular}{lcc}
\toprule
\textbf{Parameter} & \textbf{Qwen2.5-32B-Instruct} & \textbf{DAPO-Qwen-32B} \\
\midrule
Max Turn & 3 & 3 \\
Training Batch Size & 128 & 1024 \\
Mini-Batch Size & 16 & 256 \\
Learning Rate (Actor) & $5 \times 10^{-7}$ & $1 \times 10^{-6}$ \\
Learning Rate (Critic) & $1 \times 10^{-5}$ & $1 \times 10^{-5}$ \\
KL Coefficient & 0.0 & 0.0 \\
Maximum Prompt Length & 4,096 & 4,096 \\
Maximum Response Length & 8,192 & 12,288 \\
Maximum Tool Response Length & 4,096 & 4,096 \\
Temperature & 1.0 & 1.0 \\
Training Epochs & 4 & 10 \\
\bottomrule
\end{NiceTabular}
\end{table}
\vspace{-1em}
\begin{table}[H]
\centering
\caption{RL Evaluation Hyperparameters for Qwen2.5-32B-Instruct and DAPO-Qwen-32B}
\label{RL_Evaluation_Hyperparameters}
\begin{NiceTabular}{lcc}
\toprule
\textbf{Parameter} & \textbf{Qwen2.5-32B-Instruct} & \textbf{DAPO-Qwen-32B} \\
\midrule
% Max Turn & 6 & 6 \\
Maximum Prompt Length & 4,096 & 4,096 \\
Maximum Response Length & 20,480 & 32,768 \\
Maximum Tool Response Length & 4,096 & 4,096 \\
Do Sample & True & True \\
Temperature for Pass@1 & 0.6 & 0.6 \\
Top\_p for Pass@1 & 0.9 & 0.9 \\
Rollout for Pass@1 & 10 & 10 \\
Temperature for Pass@k & 0.6 & 0.6 \\
Top\_p for Pass@k & 0.9 & 0.9 \\
Rollout for Pass@k & 128 & 128 \\
\bottomrule
\end{NiceTabular}
\end{table}

\subsection{Case Study of ReVeal}
Table \ref{case_study} shows a detailed case of ReVeal.

\begin{table}%[H]
    \caption{Case Study of ReVeal Performing Iterative Generation and Verification}
    \centering
    % \scriptsize
    \small
    \begin{tabular}{@{}p{\textwidth}@{}} 
        \toprule 
        \textbf{Question:} \\
        The Two-dimensional kingdom has a regular army of $n$ people... The Two-dimensional kingdom has $m$ vests at its disposal, the $j$-th vest's size equals $b^j$. Help mobilize the Two-dimensional kingdom's army: equip with vests as many soldiers as possible. \\
        \midrule 
        \textbf{Response}: \\
        \textbf{<generation-think>} \\
        
The problem can be solved using a dual circulation.  \\
Here's a step-by-step approach to the solution: \\
... \\
Wait, I can use a greedy approach to solve this problem. Let me try the greedy algorithm by following steps: \\
... \\
\textbf{</generation-think>} \\

\textbf{<generation-answer>} \\
\textasciigrave\textasciigrave\textasciigrave python \\
def equip\_soldiers(n, m, x, y, a, b): \\
\qquad ... \\
\textasciigrave\textasciigrave\textasciigrave\\
\textbf{</generation-answer>} \\

\textbf{<verification-think>} \\
\textcolor{red}{Now, we will verify the code with the provided test cases.} \\
\textcolor{red}{\quad - Example Input 1 should result in matching 2 pairs: (Soldier at index 1 with Vest at index 1) and (Soldier at index 3 with Vest at index 2). There is no possible matching for the remaining soldiers with the available vests within the given range.} \\
% \textcolor{red}{...} \\
\textcolor{red}{Wait, I could add an extra test case for verification: more vests than soldiers...} \\
\textbf{</verification-think>} \\

\textbf{<verification-answer>} \\
- Input: \\
\textasciigrave\textasciigrave\textasciigrave 1 3 5 \textasciigrave\textasciigrave\textasciigrave \\
- Expected Output: \\
\textasciigrave\textasciigrave\textasciigrave 3 2 \textasciigrave\textasciigrave\textasciigrave \\
... \\
\textbf{</verification-answer>} \\
\textbf{<tool-feedback>} \\
... \\
- Actual Output: \\
\textasciigrave\textasciigrave\textasciigrave 3 1 \textasciigrave\textasciigrave\textasciigrave \\
\textcolor{red}{- Judgement:} \\
\textcolor{red}{Failed} \\
... \\

\textbf{</tool-feedback>} \\\\
\textbf{<generation-think>} \\
\textcolor{red}{The initial code provided did not successfully handle the actual output as expected. There is a discrepancy where the vest index tracked in the loop is not correctly being reset or updated...} \\

\textcolor{red}{Alternatively, I should create a list that keeps track of which vests have already been used and adjust the code logic to ensure each vest is used only once. Here is a revised version of the code:} \\
... \\
\textbf{</verification-answer>} \\
\textbf{<tool-feedback>} \\
... \\
- Actual Output: \\
\textasciigrave\textasciigrave\textasciigrave 3 2 \textasciigrave\textasciigrave\textasciigrave \\
\textcolor{red}{- Judgement:} \\
\textcolor{red}{Passed} \\
... \\
\textbf{</tool-feedback>}

        \\
        \bottomrule 
    \end{tabular}
    \label{case_study}
\end{table}

\subsection{Prompt Templates} 
Table \ref{Prompt_Table} shows the comparison between the commonly used Think-Answer prompt and our Generation-Verification Prompt. Our prompt guides the model to continuously alternate between generation and verification until the correct answer is obtained. Additionally, to enable the extraction of code generated by the model for providing accurate training rewards, we instruct the model to enclose the code within python blocks.

\subsection{Templates Used for Tool Feedback}
Table \ref{feedback_mapping} shows the mapping between execution results and hint templates: (1) for test cases that are verified as successful, we give a [Passed] signal in the judgement area; (2) for test cases that are verified as failed, we give a [Failed] signal in the judgement area; (3) for test cases that are verified as wrong, we give a clear feedback of [Wrong test case] for individual failures, or [No correct test cases generated] if all test cases are invalid; (4) for format error, we will give the feedback of formatting instructions to guide correct generation.

\begin{table}%[H]
  \caption{Comparison Between Think-Answer Prompt and ReVeal Prompt}
  \label{Prompt_Table}
  % \small
  \scriptsize
  \renewcommand\cellalign{lt} % 左顶对齐
  \renewcommand\cellgape{\gape[t]} % 保持垂直对齐
  \newcolumntype{L}{>{\RaggedRight\arraybackslash}X}
  \centering
\begin{NiceTabular}{p{5cm} | p{8cm}}
    \toprule
    \rowcolor{gray!15}\textbf{Think-Answer Prompt} & \textbf{ReVeal Prompt}  \\
    \midrule
    \makecell[lp{1\linewidth}]{\textbf{system} \\ You are Qwen, created by Alibaba Cloud. You are a helpful assistant. \\ \textbf{user} \\ \{question\} \\\\ First think about the reasoning process in the mind and then provides the user with the answer. The reasoning process and answer are enclosed within <think> </think> and <answer> </answer> tags, respectively, i.e., <think> reasoning process here </think> <answer> answer here </answer>. \\ Enclose your code within delimiters as follows. \\ \textasciigrave\textasciigrave\textasciigrave python \\ YOUR CODE HERE \\ \textasciigrave\textasciigrave\textasciigrave \\ \textbf{assistant} \\ } & \makecell[lp{1\linewidth}]{ \textbf{system} \\ You are Qwen, created by Alibaba Cloud. You are a helpful assistant. \\ \textbf{user} \\ \{question\} \\\\ First think through the reasoning process and write Python \\code to solve the problem, enclose your reasoning process in <generation-think> </generation-think> and present the code in \\ \textasciigrave\textasciigrave\textasciigrave python \\ Your code \\ \textasciigrave\textasciigrave\textasciigrave \\ within <generation-answer> </generation-answer> tags. After that, verify your code by generating test cases: \\\\ 1. Extract sample test cases if the problem description includes them. When necessary, generate a small number of additional test cases to validate the correctness of the generated code. \\ 2. Enclose your reasoning process in <verification-think> </verification-think> tags and enclose the final test cases and your verification conclusion within <verification-answer> </verification-answer> tags and wrap each test case using the following format: \\ - Input: \\ \textasciigrave\textasciigrave\textasciigrave \\ testcase input \\ \textasciigrave\textasciigrave\textasciigrave \\ - Expected Output: \\ \textasciigrave\textasciigrave\textasciigrave \\ expected testcase output \\ \textasciigrave\textasciigrave\textasciigrave \\ 3. Note that for "Use Call-Based format" questions, the testcase input should use a function call format, e.g., fn\_name(12, 12, 12). \\ \textbf{assistant} \\ }     \\
    \bottomrule
  \end{NiceTabular}
\end{table}

\begin{table}%[H]
    \caption{Tool Feedback Templates for Different Execution Result Types.}
    \label{feedback_mapping}
    \small
    \centering
    \begin{NiceTabular}{p{3cm} p{10cm}}
    \toprule
    \textbf{Execution Results} & \textbf{Feedback} \\
    % \midrule
     \rowcolor{gray!5}Success Test cases &  \makecell[l]{- Input: \\ \{input\} \\\\ - Expected Output: \\ \{expected output\} \\\\ - Actual Output: \\ \{actual output\} \\\\ - Judgement \\ Passed\\\\} \\
     % \midrule
     \rowcolor{gray!15}Failed Test cases &  \makecell[l]{- Input: \\ \{input\} \\\\ - Expected Output: \\ \{expected output\} \\\\ - Actual Output: \\ \{actual output\} \\\\ - Judgement \\ Failed \\\\ - Failed Reason \\ \{failed reason\}\\\\} \\
     % \midrule
     \rowcolor{gray!5}Wrong Test Cases &  \makecell[l]{- Input: \\ \{input\} \\\\ - Expected Output: \\ \{expected output\} \\\\ - Actual Output: \\ \{actual output\} \\\\ - Judgement \\ Wrong test case. \\\\ No correct test cases are generated.\\\\}  \\
     % \midrule
     \rowcolor{gray!15}Error Format &  \makecell[lp{1\linewidth}]{No valid code because of the incorrect format. Write Python code again, and present the code in \\\textasciigrave\textasciigrave\textasciigrave python\\Your code\\\textasciigrave\textasciigrave\textasciigrave\\ within <generation-answer> </generation-answer> tags. After that, verify your code by generating test cases: \\1. Extract sample test cases if the problem description includes them... \\2. Wrap each test case using the following format: \\- Input:\\\textasciigrave\textasciigrave\textasciigrave\\testcase input\\\textasciigrave\textasciigrave\textasciigrave\\- Expected Output:\\\textasciigrave\textasciigrave\textasciigrave\\expected testcase output\\\textasciigrave\textasciigrave\textasciigrave} \\
    \bottomrule
    \end{NiceTabular}
\end{table}



\section{The Effectiveness of Short-Term Memory} 
\label{app: memory}
To ensure context understanding in long interaction loops, we adopt a short-term memory component, which implements a short-term memory mechanism that leverages in-context learning to store recent generation-verification loops within the contextual window. The system maintains a rolling history of the most recent interactions, including code generations, test verifications, and tool feedback. Critical information such as successful patterns, error types, and effective test structures are preserved in complete formats. This structured memory representation enables the model to quickly identify relevant patterns from recent history and build upon previous attempts, leading to faster convergence and reduced redundant exploration in the solution space.

We compare models with and without memory integration. The baseline model without memory provides complete historical information from all previous turns directly to the model, maintaining full contextual details throughout the interaction sequence. We verified the impact of ReVeal on Pass@1 on LiveCodeBench with and without memory. Test results indicate that introducing short-term memory does not cause a decline in Pass@1 (w/o memory 38.2\% vs. w/ memory 38.3\% at turn 15), and may even yield a slight performance boost. This sustained improvement capability highlights the memory mechanism's effectiveness in enabling continuous learning and adaptation within computational constraints.

\section{More experiments on Qwen2.5-32B-Instruct}
\label{sec:qwen32b_more}

\paragraph{Main Results.}
% To validate effectiveness and scalability across models, we evaluate ReVeal on another base model, Qwen2.5-32B-Instruct. Table~\ref{table:main3} shows ReVeal outperforms single-turn RL baseline by 4.1\%, which demonstrates the effectiveness of ReVeal. TAPO with joint verifiable rewards yields \textbf{38.0\%} vs.\ \textbf{37.1\%} Pass@1 for outcome-only (+0.9), while maintaining a higher $\Delta\uparrow$ and near-zero $\Delta\downarrow$. Results demonstrate ReVeal remains effective across models of varying capability.
To further validate the effectiveness and scalability of ReVeal across different base models, we additionally conduct experiments on Qwen2.5-32B-Instruct. As shown in Table~\ref{table:main3}, ReVeal achieves a 4.1\% improvement in Pass@1 over the single-turn RL baseline. Moreover, using TAPO with joint verifiable rewards yields higher Pass@1 compared to ReVeal with outcome-only reward, while maintaining a higher $\Delta\uparrow$ and near-zero $\Delta\downarrow$. These results demonstrate that ReVeal remains effective and stable across backbones of varying capability.

\begin{figure*}[h]
    \centering
    \includegraphics[width=\textwidth]{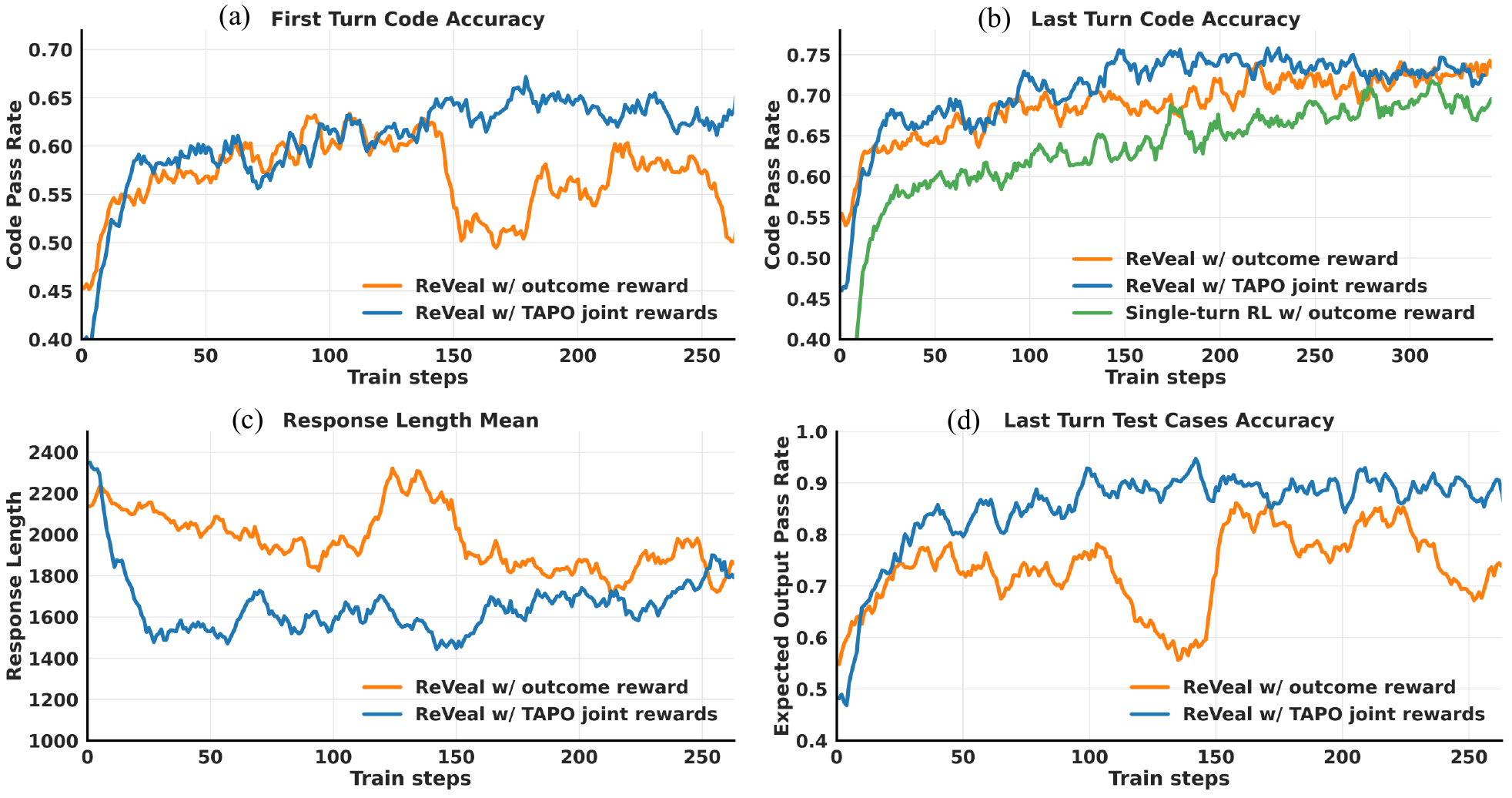}  
    \vskip -0.1in
    \caption{Comparison of code accuracy, test case accuracy, and response length across training for ReVeal (Qwen2.5-32B-Instruct) with turn-level rewards, ReVeal with outcome-only rewards, and single-turn RL without tool integration.} 
    \label{fig:experiment2}
    \vskip -0.1in
\end{figure*}

\paragraph{Training Curves.}
Figure~\ref{fig:experiment2} shows concurrent improvements in both code accuracy and test-case accuracy when using TAPO with joint rewards, indicating that explicitly optimizing self-verification leads to more reliable verification and drives stronger multi-turn refinement, echoing our main findings on DAPO-Qwen-32B.

\paragraph{Significance.}
To demonstrate that the performance improvements are statistically significant, we repeated the experiment 8 times and reported the $mean\pm std$ in Table~\ref{table:main4}, confirming the significance of the gains.

% \paragraph{Takeaway across models.}
% Together with the main-text DAPO results, these findings demonstrate that \method{} consistently enhances performance across models of varying capability; self-verification remains a reliable driver of deeper test-time scaling, and stronger backbones unlock larger headroom, while on \texttt{Qwen2.5-32B-Instruct} we already observe clear benefits.

\begin{table}[h]
    \centering
    \caption{Performance comparison of ReVeal (Qwen2.5-32B-Instruct) with baseline methods on LiveCodeBench. Pass@1 indicates the success rate; $\Delta_{\uparrow}$ and $\Delta_{\downarrow}$ represent the percentages of incorrect solutions corrected and correct solutions degraded after revision, respectively.}
    \begin{NiceTabularX}{\textwidth}{lX[c]X[c]X[c]}
    \toprule
    % Model column spans 2 rows. LiveCodeBench spans the 3 X-columns.
    \textbf{Model} & \multicolumn{3}{c}{\textbf{LiveCodeBench V5}} \\
    \cmidrule(lr){2-4} % Rule under LiveCodeBench, spanning columns 2 to 4
    % Sub-headers for the 3 data columns
    & \textbf{Pass@1} & $\Delta_{\uparrow}$ & $\Delta_{\downarrow}$ \\
    \midrule
    \multicolumn{4}{c}{\rowcolor{gray!15}\textit{Existing Baselines}} \\
    Qwen2.5-32B-Instruct            & 26.6  & -     & -     \\
    DAPO-Qwen-32B & 29.6 & - & - \\
    Qwen2.5-Coder-32B-Instruct      & 30.5    & -     & -     \\
    \phantom{1ex}w/ critic$\times$5 Qwen2.5-Coder & 29.6 & 2.14   & 3.04  \\
    \phantom{1ex}w/ critic$\times$5 GPT-4o         & 32.9 & 4.82  & 2.50  \\
    \phantom{1ex}w/ critic$\times$5 CTRL           & 33.4 & 3.75  & 0.89  \\
    % DeepSeek-R1-Zero-Qwen-32B & 40.2 & - & -  \\
    \midrule
    \multicolumn{4}{c}{\rowcolor{gray!15}\textit{RL based on Qwen2.5-32B-Instruct}} \\
    Single-turn RL & 33.9  & -   & -   \\
    ReVeal$\times$6 & \textbf{38.0}    & \textbf{3.41}   & \textbf{0.0}   \\
    \midrule
    % \multicolumn{4}{c}{\rowcolor{gray!15}\textit{RL based on DAPO-Qwen2.5-32B}} \\
    % Single-turn RL w/ outcome reward& 36.5  & -   & -   \\
    % ReVeal$\times$19 w/ turn-level reward& \textbf{42.4}    & \textbf{10.56}   & \textbf{0.18}   \\
    % \midrule
    \multicolumn{4}{c}{\rowcolor{gray!15}\textit{Ablation Study: TAPO with Joint Verifiable Rewards}} \\
    ReVeal$\times$6 w/ outcome reward & 37.1    & 2.98   &  0.0   \\
    ReVeal$\times$6 w/ TAPO with joint rewards & 38.0    & 3.41   & 0.0   \\
    \bottomrule
    \end{NiceTabularX}
    \label{table:main3}
\end{table}

\begin{table}[h]
    \centering
    \caption{Significance test of ReVeal (Qwen2.5-32B-Instruct) on LiveCodeBench V5. $mean\pm std$ indicates the average code Pass@1 from 8 repeated experiments.}
    \begin{NiceTabularX}{\textwidth}{lX[c]}
    \toprule
    % Model column spans 2 rows. LiveCodeBench spans the 3 X-columns.
    \textbf{Model} & \multicolumn{1}{c}{\textbf{LiveCodeBench V5}} \\
    \cmidrule(lr){2-2} % Rule under LiveCodeBench, spanning columns 2 to 4
    % Sub-headers for the 3 data columns
    & $mean\pm std$ \\
    \midrule
    % ReVeal$\times$3 turn w/ turn-level reward            &  37.37 $\pm$ 0.18      \\
    ReVeal$\times$6 w/ outcome reward           &  37.09 $\pm$ 0.33      \\
    ReVeal$\times$6 w/ TAPO with joint rewards            &  38.02 $\pm$ 0.31      \\
    \bottomrule
    \end{NiceTabularX}
    \label{table:main4}
\end{table}

\section{Compare with prompt-only ReVeal}
\label{promptsec}
To verify that the performance improvements originate from explicit optimization rather than mere tool use and prompting, we evaluate a \emph{prompt-only ReVeal} variant on Qwen2.5-32B-Instruct that keeps the same ReVeal prompt and the same code-execution tool, but performs no RL training. As shown in Table~\ref{table:main2}, this prompt-only setup fails to sustain effective deep multi-turn refinement, indicating that simply changing prompts or enabling tools is insufficient. In contrast, ReVeal with RL-trained verification delivers consistent gains and markedly higher verification reliability; for example, test-case accuracy improves from around \(50\%\) to about \(88\%\) after training (Figure~\ref{fig:experiment}(d)), confirming that the gains chiefly stem from explicitly optimized self-verification rather than prompting or tool use alone.

% Reflexion \citep{shinn2023reflexion} uses GPT-4 to generate code and tests, call tools for feedback, and refine code on easier benchmarks like HumanEval and MBPP, yet still suffers from inaccurate tests (causing 59\% false negatives and 16\% false positives), which can even hurt performance. Our work targets harder code competition problems and explicitly optimizes self-verification — a key factor for LongCoT success. As shown in Figure~\ref{fig:experiment}(d), test accuracy improves from approximately 50\% to 88\% after ReVeal RL training.

% As shown in Table~\ref{table:main2}, we tested ReAct-style prompting on Qwen2.5-32B-Instruct with the same tool. Without RL-trained verification, it failed to support effective deep multi-turn refinement, highlighting the necessity of ReVeal training for stronger generation as well as enhanced verification and refinement capabilities to enable deeper scaling.

\begin{table}[H]
    \centering
    \caption{Performance comparison of ReVeal with prompt-only variant on Qwen2.5-32B-Instruct. Pass@1 indicates the success rate; $\Delta_{\uparrow}$ and $\Delta_{\downarrow}$ represent the percentages of incorrect solutions corrected and correct solutions degraded after revision, respectively.}
    \begin{NiceTabularX}{\textwidth}{lX[c]X[c]X[c]}
    \toprule
    % Model column spans 2 rows. LiveCodeBench spans the 3 X-columns.
    \textbf{Model} & \multicolumn{3}{c}{\textbf{LiveCodeBench V5}} \\
    \cmidrule(lr){2-4} % Rule under LiveCodeBench, spanning columns 2 to 4
    % Sub-headers for the 3 data columns
    & \textbf{Pass@1} & $\Delta_{\uparrow}$ & $\Delta_{\downarrow}$ \\
    \midrule
    ReVeal Prompting w/o Training           &    &       &      \\
    \phantom{1ex}$\times$1 turn & 26.3 & - & - \\
    \phantom{1ex}$\times$3 turn & 27.5 & - & - \\
    \phantom{1ex}$\times$6 turn & 27.4 & 2.40 & 1.31 \\
    ReVeal &      &     &     \\
    \phantom{1ex}$\times$1 turn & 35.7 & - & - \\
    \phantom{1ex}$\times$3 turn & 37.5 & - & - \\
    \phantom{1ex}$\times$6 turn & 38.0 & 3.41 & 0.0 \\
    \bottomrule
    \end{NiceTabularX}
    \label{table:main2}
\end{table}

% As shown in Figure~\ref{fig:experiment3}, we compare models with and without memory integration. The baseline model without memory provides complete historical information from all previous turns directly to the model, maintaining full contextual details throughout the interaction sequence. However, experimental results reveal critical limitations of this approach. While both configurations show similar performance improvements in early interaction rounds, the no-memory baseline exhibits performance saturation at deeper interaction turns due to output length constraints imposed by the model's context window. In contrast, the memory-augmented system demonstrates sustained performance gains throughout extended interaction sequences. Specifically, at 20+ interaction turns, the memory-enabled model continues to achieve incremental improvements, while the baseline's performance plateaus and occasionally degrades due to context overflow issues. This sustained improvement capability highlights the memory mechanism's effectiveness in enabling continuous learning and adaptation within computational constraints.

% \begin{figure}[t]
%     \centering
%     \includegraphics[width=0.6\textwidth]{figs/fig11.pdf}  
%     \vskip -0.1in
%     \caption{Comparison of ReVeal with Memory and without Memory.} 
%     \label{fig:experiment3}
%     \vskip -0.1in
% \end{figure}
% \appendix
% \section{Appendix}

\end{document}

%% file: math_commands.tex
\usepackage{amsmath,amsfonts,bm}

\def\eqref#1{equation~\ref{#1}}
\def\1{\bm{1}}

\DeclareMathAlphabet{\mathsfit}{\encodingdefault}{\sfdefault}{m}{sl}
\SetMathAlphabet{\mathsfit}{bold}{\encodingdefault}{\sfdefault}{bx}{n}